\documentclass[aps,pra,reprint,showpacs,floatfix,superscriptaddress]{revtex4-2}

\usepackage{graphicx}
\usepackage{dcolumn}
\usepackage{bm}
\usepackage{placeins}
\usepackage[percent]{overpic}
\usepackage{xcolor}
\usepackage{xcolor}
\usepackage{amsmath}
\usepackage{multirow}
\usepackage{graphicx}
\usepackage{braket}
\usepackage[percent]{overpic}
\usepackage{tikz}
\usepackage{amsfonts}
\usepackage{mathtools}
\begin{document}

\preprint{APS/123-QED}

\title{Quantum-enhanced estimation of stimulated Raman optical activity}

\author{Mahadeva Chanda Durjoy}
\affiliation{Department of Physics and Astronomy, Texas A\&M University, College Station, Texas 77843, USA}
\affiliation{Institute for Quantum Science and Engineering, Texas A\&M University, College Station, Texas 77843, USA}
\author{Girish S. Agarwal}
\affiliation{Department of Physics and Astronomy, Texas A\&M University, College Station, Texas 77843, USA}
\affiliation{Institute for Quantum Science and Engineering, Texas A\&M University, College Station, Texas 77843, USA}
\affiliation{Department of Biological and Agricultural Engineering, Texas A\&M University, College Station, Texas 77843, USA}

\begin{abstract}

In recent times there has been growing interest in Raman optical activity (ROA) for its label free detection of absolute configuration, conformation, and stereochemical structure in chiral biosamples and drug molecules. Since ROA signals are generally small, techniques such as stimulation by a probe beam can be used to enhance the signal strength. However, with a classical probe, the measurement precision is still fundamentally limited by its shot noise. To solve this problem we propose the use of two-mode squeezed vacuum and show that it can achieve sub-shot noise limited measurement sensitivity. Using quantum estimation theory, we derived the quantum Fisher information and the quantum Cram\'er-Rao bound (QCRB) for stimulated ROA measurement to quantify the precision enhancement. This improvement comes from photon-number correlations which suppress the intensity fluctuation common to both modes. We further show that balanced detection of the output intensity difference is a practical measurement scheme that approaches the QCRB and becomes optimal in the small-chirality limit. This opens a promising path toward more sensitive Raman chiroptical spectroscopy of weak and photosensitive samples.

\end{abstract}

\maketitle

\section{Introduction} 
Improving measurement precision is the central goal of quantum metrology, which takes advantage of quantum effects using quantum resources exhibiting quantum features such as entanglement, squeezing and non-classical correlations to achieve sensitivities beyond classical limits~\cite{caves1981,yurke1986,giovanetti2004}. These effects have been exploited in many high precision measurements, and used in applications like atomic clocks~\cite{pezze2018,pedrozo2020}, phase shift measurement~\cite{xiao1987,holland1993}, atomic force microscopy~\cite{pooser2020}, gravitational wave detection~\cite{grote2013,meylahn2022,jia2024}, spectroscopy and imaging~\cite{andrade2020,casacio2021,li2022}. Raman optical activity (ROA) is a notable example where the measured signal is extremely small, but it carries highly valuable physical information~\cite{Barron2004ofage,barron2015}. This makes high-precision measurement essential for reliable ROA detection. ROA is an inelastic light-scattering process in which an incident pump beam interacts with a molecule to excite it to a vibrational state and produce a scattered Stokes beam at a lower frequency. For chiral molecules, the intensities of the left circularly polarized(LCP) and right circularly polarized(RCP) components of the scattered light are slightly different, and in an ROA experiments this intensity difference is the measured quantity~\cite{barron2004book,barron2006}. This difference is typically only $10^{-3}$ to $10^{-5}$ of the total Raman signal~\cite{abdali2008}, which makes its precision measurement highly challenging. Although the ROA signal is extremely small, it is very informative because it originates from several light–matter interaction mechanisms, including contributions from electric-dipole, magnetic-dipole, and electric-quadrupole.~\cite{barron2004book,parchansky2014}

The sensitivity of ROA to these higher order light–matter interactions enables it to probe molecular chirality~\cite{barron2007} in ways not accessible to vibrational spectroscopies like infrared spectroscopy or conventional Raman spectroscopy. Consequently, ROA spectra provide information about absolute configuration~\cite{polavarapu2002,haesler2007}, conformations~\cite{barron2015,schrenkova2023} and structure of chiral molecules~\cite{Barron2004ofage,profant2026}. This structural sensitivity has made ROA particularly valuable for studying the secondary structure of protein~\cite{blanch2003,zhu2005protein}. It is also used to study other biomolecular systems such as, nucleic acids~\cite{blanch2003}, peptide~\cite{zhu2008} and carbohydrates~\cite{zhu2006all,barron2015}, where subtle changes in three dimensional structure are closely linked to biological function. It has also found important applications in pharmaceutical research~\cite{barron2015} for determining stereochemistry~\cite{bogaerts2021} and enantiomer discrimination~\cite{tian2025} of chiral drug molecules, as well as in the characterization of other chiral molecular systems in chemistry and materials science~\cite{sun2013,barron2015,li2025}. Thus, despite being a weak, ROA signal remains a powerful probe of molecular structure that provides detailed geometric and chiral information.

The primary drawback of Raman spectroscopy is the intrinsic weakness of its signal strength. Thus, conventional ROA measurements often require tens of hours to even days~\cite{blanch2001} for the signal to achieve adequate signal-to-noise ratios. Increasing pump beam intensity to improve signal strength is also not always an option for sensitive molecular or biological samples due to photobleaching or phototoxicity~\cite{fu2006photodamage,zhang2017damage,zhang2022phototoxic}. In conventional Raman spectroscopy this limitation is mitigated by employing stimulated Raman processes, where two optical fields (a pump and a Stokes beam) coherently drive vibrational transitions and generate an amplified Raman response~\cite{jones2019,li2021srs,min2025,gao2025}. Such schemes can enhance the detected signal~\cite{li2021srs} and reduce acquisition times. In a stimulated ROA measurement, the Stokes field probing the chiral sample experiences slightly different stimulated Raman gain in the LCP and RCP. Then the ROA signal is obtained at the output by measuring the small intensity difference between the light of these two circular polarization. Since this chiral gain difference is masked by a much relatively larger gain common to both circular polarization modes, the measurement is strongly affected by shot noise associated with the total intensity of the signal, and this can offset the advantage gained from signal amplification. This trade-off between signal enhancement and noise ultimately constrains the sensitivity achievable with classical light, motivating the exploration of quantum optical probes that can modify the fundamental noise properties of the measurement stemming from the photon number fluctuation of the probe, and potentially enable more sensitive detection of weak chiral Raman signals.

The use of squeezed light to surpass shot-noise-limited sensitivity in stimulated Raman scattering (SRS) has been investigated theoretically~\cite{anatoly2021,schlawin2026}, identifying nonclassical states as advantageous resources. Experimentally, sub-shot-noise sensitivity in SRS has been demonstrated using amplitude-squeezed light~\cite{andrade2020,casacio2021}. In a closely related chiroptical technique, namely circular dichroism (CD) sensing, theoretical studies have similarly shown that nonclassical probes such as two-mode squeezed vacuum and bright squeezed states can provide a quantum advantage in sensitivity~\cite{ioannou2021,wang2021,belsley2022}. To the best of our knowledge, the use of quantum probe states in stimulated Raman optical activity (ROA) has not been systematically investigated. Motivated by this limitation, we develop a quantum-estimation framework for stimulated ROA in which the common Raman gain $G_0$ and the chirality-induced gain difference $G_\chi$ are treated as jointly unknown parameters. Within this multiparameter setting, we derive the quantum Cram\'er--Rao bound for the coherent probes which sets the standard quantum limit for estimating $G_\chi$. We then show that a two-mode squeezed vacuum (TMSV) probe yields a lower estimation error at the same mean photon number, with the largest advantage in the weak-gain regime relevant to stimulated ROA. In addition, we identify balanced detection of the output intensity difference as a practical measurement scheme and show that it approaches the quantum limit, becoming optimal in the small-chirality regime. These results establish quantum-correlated light as a promising resource for improving ROA sensitivity without increasing optical power.

\section{Theoretical Framework}

In stimulated Raman optical activity (ROA), a pump beam and a Stokes probe beam of lower frequency are used to measure the chiral properties of a sample. When the frequency difference between these two beams matches with a molecular vibrational transition, energy is transferred coherently from the pump to the probe, producing a stimulated Raman gain on the probe beam~\cite{min2025}. In a chiral medium this gain can be polarization dependent, and different circular polarization components may experience slightly different gain. This gain difference can be found from the intensity difference of LCP and RCP of the probe beam. However, other mechanisms that can also contribute to the difference in intensity must be excluded, so that the measured intensity difference can be attributed solely to difference in stimulated Raman gain. In general the difference in intensity can also arise due to polarization mixing between LCP and RCP, residual birefringence or circular dichroism.

In this work, for practical purposes we neglect polarization mixing and allow the two circular polarization modes to evolve independently. This approximation is well justified for the many aqueous samples, where chiral molecules are typically isotropic or randomly oriented in solution~\cite{barron2000}. For such samples, the medium is rotationally invariant and linear birefringence averages to zero. Additionally we assume that the frequency of pump and probe are chosen such that the absorption is weak in that spectral region. For many biosamples the electronic absorption and circular-dichroism primarily occurs at the ultraviolet region~\cite{ranjbar2009,rogers2019}. Thus, for aqueous solution, visible spectrum can be used due to low absorbance of light. This is not a limitation of stimulated ROA, since the Raman resonance condition is determined by the pump–probe frequency difference. And it can be tuned to a molecular vibrational transition even when both optical carrier frequencies lie within a low-absorbance window. Under these conditions, the contribution of ordinary circular dichroism in the measured left- and right-circularly polarized intensity difference is negligible. Finally, we work in the undepleted-pump and weak-gain regime, as is appropriate for the systems considered 

\begin{figure}
    \centering
    \begin{overpic}[width=1\linewidth]{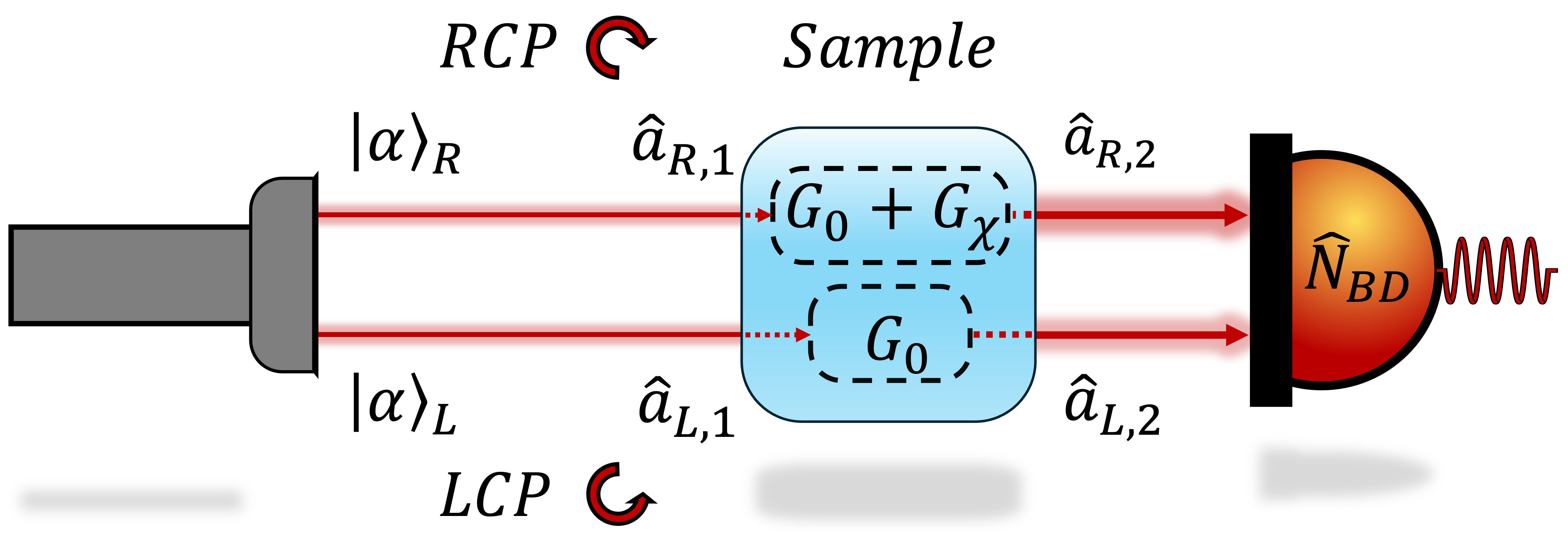}
    \end{overpic}
    \caption{Schematics for stimulated Raman optical activity sensing: right and left circular polarization mode of coherent Stokes beam undergoing intensity gain by a factor of $G_0+G_\chi$ and $G_0$ respectively upon passing through the sample (The pump used on the sample for gain stimulation is not shown in diagram). Intensity difference at the output is measured to find the gain difference $G_\chi$.}
    \label{fig:Laser}
\end{figure}

Under these assumptions, we model the stimulated Raman interaction for the two circular polarization modes of the probe with the sample as two independent phase-insensitive weak optical amplifier channels. Since the chiral contribution is expected to be small, it is convenient to separate the gain into a circular polarization independent common gain \(G_0\) and a small chirality induced gain difference \(G_\chi\), which may be positive or negative depending on the sign of the chiral Raman response in the spectral region probed~\cite{Barron2004ofage}. Choosing the left-circularly polarized channel as a reference, we write the gain for Stokes beam of LCP and RCP as $G_L=G_0$ and $G_R=G_0+G_\chi$ respectively. This is schematically represented in Fig.~\ref{fig:Laser}. The master equation for quantum optical amplification can be used to derive the quantum Langevin equation, and its solution is discussed in Eqs.(10.7)-(10.12) of the textbook~\cite{Agarwal_2012,raymer81,raymer85}. We use the corresponding input--output relations for sample induced gain in Stokes probe of each of the two circular polarization channels, 
\begin{equation}{
    \begin{aligned}
    \hat a_{\mathrm{R,2}}&=
    \sqrt{G_0+G_\chi}\,\hat a_{\mathrm{R,1}}
    +
    \sqrt{G_0+G_\chi-1}\,\hat b_{\mathrm{R}}^{\dagger},\\
    \hat a_{\mathrm{L,2}}&=
    \sqrt{G_0}\,\hat a_{\mathrm{L,1}}
    +
    \sqrt{G_0-1}\,\hat b_{\mathrm{L}}^{\dagger},
    \end{aligned}}
    \label{ipop}
\end{equation}
where
$[\hat a_{\sigma,i}^{\vphantom{\dagger}},\hat a_{\sigma',j}^{\dagger}]
=\delta_{\sigma\sigma'}\delta_{ij}$
and
$[\hat b_{\sigma}^{\vphantom{\dagger}},\hat b_{\sigma'}^{\dagger}]
=\delta_{\sigma\sigma'}$. Here \(\hat a_{\sigma,1}\) and \(\hat a_{\sigma,\mathrm{2}}\) are the probe annihilation operators at the input and output of the sample \(i,j\in\{1,2\}\), respectively, for the circular polarization mode \(\sigma\in\{L,R\}\), shown in Fig.~\ref{fig:Laser}. While \(\hat b_\sigma\) is the associated environmental mode for the amplification process. It preserves the bosonic commutation relation at the output and accounts for the added noise intrinsic to the amplification process. In the appendix C we have derived the input-output relation for the Raman gain in each polarization channel using the effective Hamiltonian for Raman process and the model for noise arising from dephasing effects.

\section{Ultimate Precision from the Quantum Fisher Information via the Quantum Cram\'er–Rao Bound}
To investigate the ultimate precision with which the chirality induced gain difference $G_\chi$ can be estimated we use quantum Fisher information(QFI). QFI quantifies how sensitively a quantum state depends on a parameter, that is, how distinguishable neighboring quantum states become under an infinitesimal change of that parameter~\cite{Braunstein1994,paris2009}. A larger quantum Fisher information therefore implies a smaller achievable estimation uncertainty. 

The proper analysis for $G_\chi$ requires local multiparameter quantum estimation theory. Estimating $G_\chi$ alone would implicitly assume that the common gain $G_0$ is known exactly. In practice, however, $G_0$ is also uncertain, and its uncertainty is of the same order as that of $G_\chi$. Treating $G_0$ as perfectly known would therefore lead to an artificially optimistic estimate of the sensitivity for $G_\chi$. For this reason, $G_\chi$ and $G_0$ must be treated as jointly unknown parameters and estimated simultaneously. For our problem, let $\boldsymbol{\epsilon}=(G_0,G_\chi)^T$ denote the set of unknown parameters encoded in the density operator $\rho_{\boldsymbol{\epsilon}}$. The corresponding lower bound on the covariance matrix of a locally unbiased estimator vector $\boldsymbol{\hat \epsilon}=(\hat G_0,\hat G_\chi)^T$ , is given by the quantum Cram\'er-Rao bound(QCRB)~\cite{paris2009},

\begin{align}
\mathrm{Cov}(\boldsymbol{\hat \epsilon}) \ge \frac{1}{\mathcal{N}_m}\left(\mathbf{F}_{Q}\right)^{-1},
\end{align}
where $\mathrm{Cov}(\boldsymbol{\hat \epsilon})$ is the covariance matrix of $\boldsymbol{\hat\epsilon}$, $\textbf{F}_Q$ is the Quantum Fisher information and $\mathcal{N}_m$ is the number of independent measurements performed, which is taken to be one for simplicity. We next show how to calculate $\textbf{F}_Q$.

In our case, the sample behaves as a weak optical amplifier. Since the input-output relation of the Stokes probe is linear in the bosonic annihilation and creation operators\eqref{ipop}, it defines a Gaussian-preserving channel, so Gaussian input states remain Gaussian after the evolution~\cite{agarwal1987}. By defining a vector of operators $ \boldsymbol{\hat A}_{\mathrm{out}}\coloneqq(\hat a_{R,2},\hat a_{L,2},\hat a^\dagger_{R,2},\hat a^\dagger_{L,2})$, the displacement vector is,
\begin{equation}
    \begin{aligned}
    d^m=\text{tr}(\hat\rho_{in}\hat {A}_{out}^m)
    \end{aligned}
    \label{disp},
\end{equation}
and the covariance matrix,
\begin{equation}
    \begin{aligned}
    \sigma^{mn}=\text{tr}(\hat\rho_{in}(\Delta\hat {A}_{out}^m\Delta\hat {A}_{out}^{n\dagger}+\Delta\hat {A}_{out}^n\Delta\hat {A}_{out}^{m\dagger})),
    \end{aligned}
    \label{covar}
\end{equation}
which makes the Gaussian-state formalism a natural framework for evaluating the quantum Fisher information.

For parameters encoded through a Gaussian-preserving evolution, the quantum Fisher information is found to be,
\begin{equation}
    (F_Q)_{ij}=\lim_{\nu \to 1}\left[\tfrac{1}{2}\mathrm{vec}[\partial_i \sigma]^\dagger M^{-1}\mathrm{vec}[\partial_j \sigma]+2\partial_i \mathbf{d}^\dagger \sigma^{-1}\partial_j \mathbf{d}\right],
    \label{QFI}
\end{equation}

where $M=\left(\nu^2 \sigma \otimes \sigma - K \otimes K\right)$ and $i,j\in\{G_0,G_\chi\}$~\cite{,gao2014,safranek2019}. In our convention, $K = \mathbb{I}_2 \oplus (-\mathbb{I}_2)$ and $\mathbb{I}_2$ is the $2\times2$ identity matrix, and vec[.] denotes vectorization of a matrix. In the following subsection we use this equation to find the expression for quantum Fisher information and the highest precision bound set by QCRB for coherent light and two mode squeezed vacuum. These expressions remain valid for both positive and negative values of the chiral gain difference $G_\chi$. And the common gain can be associated with any polarization mode, we have chosen for illustration the common gain to be associated with the left polarization and considered the positive values of $G_\chi$.

\subsection*{The Standard Quantum Limit set by coherent light}
To establish the classical benchmark for the simultaneous estimation of the common gain $G_0$ and the chirality induced gain difference $G_\chi$, we consider a probe composed of coherent states in the right- and left-circularly polarized modes (as shown in Fig.~\ref{fig:Laser}),
\begin{equation}
    \ket{\psi^{\mathrm{Coh}}}=\ket{\alpha}_R\otimes\ket{\alpha}_L,
\end{equation}
Since the estimation problem is phase insensitive, the phase of the coherent state is irrelevant and can be taken to be zero, thus we use $\alpha\in\mathbb{R}$. With this, each mode contains the same mean photon number $\bar n=\alpha^2$. And for a valid comparison we later use the same photon number in each channel of TMSV.

Because the sample implements a Gaussian-preserving amplification channel, the output state remains Gaussian and is fully specified by its displacement vector and covariance matrix. From the input-output relations in Eq.~\eqref{ipop} and the expressions in Eqs.~\eqref{disp} and \eqref{covar}, the output displacement vector is given by
\begin{equation}
\mathbf{d}_{\mathrm{coh}}=
\begin{pmatrix}
\alpha\sqrt{G_0+G_\chi}\\
\alpha\sqrt{G_0}\\
\alpha\sqrt{G_0+G_\chi}\\
\alpha\sqrt{G_0}
\end{pmatrix},
\end{equation}
and the covariance matrix is
\begin{equation}
\sigma_{\mathrm{coh}}=
\begin{pmatrix}
\sigma_R & 0 & 0 & 0\\
0 & \sigma_L & 0 & 0\\
0 & 0 & \sigma_R & 0\\
0 & 0 & 0 & \sigma_L
\end{pmatrix},
\end{equation}
where, $\sigma_R=2(G_0+G_\chi)-1$ and $\sigma_L=2G_0-1$. 
The covariance matrix is diagonal because the two modes of the coherent probe are uncorrelated and they evolve independently through the sample.

The displacement vector and the covariance matrix can now be used along with Eq.~\eqref{QFI} to find the quantum Fisher information matrix for gain estimation with coherent state
\begin{equation}
\mathbf{F}^{\mathrm{coh}}_{Q}=
\begin{pmatrix}
A+B & A\\
A & A
\end{pmatrix},
\end{equation}
with
\begin{align}
A &=\frac{4}{\sigma_R^2-1}+ \frac{\alpha^2}{\sigma_R(G_0+G_\chi)},\\[4pt]
B &= \frac{4}{\sigma_L^2-1}+\frac{\alpha^2}{\sigma_L G_0},
\end{align}
where the QFIM is written in the ordered parameter basis $\boldsymbol{\epsilon}=(G_0,G_\chi)$.
The terms proportional to $\alpha^2$ quantify the enhancement of the QFI due to stimulated emission induced by the coherent probe. In contrast, the $\alpha$-independent terms represent the contribution from spontaneous emission, which remains present even with vacuum as the input. In the same basis the inverse of QFIM is
\begin{equation}
\left(\mathbf{F}^{\mathrm{coh}}_{Q}\right)^{-1}
=
\frac{1}{AB}
\begin{pmatrix}
A & -A\\
-A & A+B
\end{pmatrix},
\end{equation}
so the multiparameter QCRB is given
\begin{align}
    (\Delta G_\chi)^{\mathrm{Coh}}_{\mathrm{QCRB}} &\ge \sqrt{\left(\mathbf{F}^{\mathrm{coh}}_{Q}\right)^{-1}_{22}}
    = \sqrt{\frac{1}{A}+\frac{1}{B}}.
    \label{Cohqcrbfull}
\end{align}
This is the standard quantum limit for coherent probe in the present multiparameter estimation problem. 

The gain coefficient of stimulated Raman signal is proportional to the spontaneous Raman scattering cross section.  Since, for most biosamples ROA is on the order of $10^{-3}$ of the total Raman
signal~\cite{barron2015} the stimulated Raman gain difference is much smaller than the actual gain. In this limit, $|G_\chi |\ll(G_0-1)$ the QCRB is approximately independent of $G_\chi$ and the Eq.\eqref{Cohqcrbfull} reduces to 
\begin{align}
(\Delta G_\chi)^{\mathrm{Coh}}_{\mathrm{QCRB}}\approx\sqrt\frac{2 (G_0-1) G_0}{\frac{\bar n}{\gamma}+1},
\label{Cohqcrb}
\end{align}
where $\gamma=\frac{G_0}{G_0-1}+1$. It is useful to write Eq.~\eqref{Cohqcrb} in terms of $\gamma$ because it will make the comparison with the QCRB of TMSV in the same limit easier, Eq.~\eqref{TMSVQCRB}. Stimulated Raman gain is only slightly greater than 1 for most realistic samples used. In the weak-gain regime, the small numerator in Eq.~\eqref{Cohqcrb} might suggest improved sensitivity; however, this apparent advantage is largely offset by the large value of $\gamma$, which reduces the effective photon-number contribution to $\bar{n}/\gamma$. As a result, the overall sensitivity is not significantly improved in this regime.

\subsection*{Beating the Standard Quantum Limit with TMSV}

\begin{figure}
    \centering
    \begin{overpic}[width=1\linewidth]{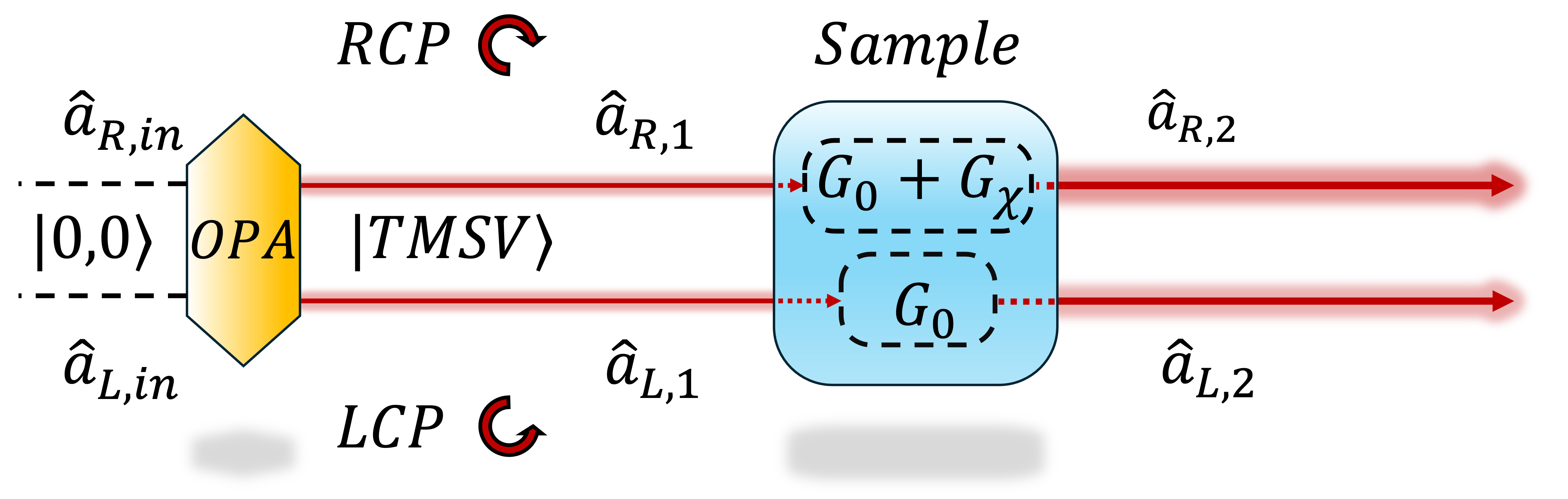}
    \end{overpic}
    \caption{Using right and left circular polarization entangled two-mode squeezed vacuum to find the QFI for gain difference ($G_\chi$) between RCP and LCP modes(The pump beam for the OPA and sample is not shown in the figure). This setup is similar to Fig~\ref{fig:Laser}, with the coherent probe replaced by TMSV. }
    \label{figure:QFI}
\end{figure}

Two-mode squeezed vacuum (TMSV) state is an entangled bipartite Gaussian state generated via spontaneous parametric downconversion in a nonlinear crystal~\cite{schumaker1985,hong1985}. Mesoscopic photon number for a single spatiotemporal mode has been demonstrated experimentally, Harder et al generated TMSV states with a mean photon number of 20 in each mode~\cite{harder16}. The entanglement in TMSV is useful primarily because it gives rise to high photon number correlation, and this property has been exploited in various applications like quantum illumination~\cite{tan2008,lopaeva2013}, quantum imaging~\cite{brambilla2008,genovese2016} and quantum sensing~\cite{meda2017}.

Under suitable phase-matching conditions, spontaneous parametric down-conversion produces photon pairs entangled in orthogonal linear polarization modes~\cite{rubin1996}. A suitable wave-plate transformation can then map these modes onto the left- and right-circular polarization basis. These modes are described by the annihilation operators $\hat a_L$ and $\hat a_R$, respectively. The resulting state obtained by applying the two-mode squeezing operator
\begin{align}
\hat S(r) = \exp\left[r\left(\hat a_{R,in}^{\dagger}\hat a_{L,in}^{\dagger} -  \hat a_{R,in}^{\vphantom{\dagger}} \hat a_{L,in}^{\vphantom{\dagger}}\right)\right],
\end{align}
to the vacuum state. Here $r$ is the squeezing strength and it can be taken to be real, since the gain measurement is phase independent.  The state is called two-mode squeezed vacuum because vacuum was used as the input state. The following calculations are most conveniently performed in the Heisenberg picture. For vacuum input, after two-mode squeezing the annihilation operators for the left- and right-circular polarization modes are
\begin{equation}
    \begin{aligned}
    \hat a_{R,1}=&\hat a_{R,in}\cosh{r}+\hat a_{L,in}^{\dagger}\sinh{r}\\
    \hat a_{L,1}=&\hat a_{L,in}\cosh{r}+\hat a_{R,in}^{\dagger}\sinh{r},
\end{aligned}
\label{a1tmsv}
\end{equation}
where the input operators obey the bosonic commutation relations $[\hat a_{\sigma,in},\hat a_{\sigma,in}^{\dagger}]=1$ with $\sigma\in{L,R}$. These transformed operators are obtained through the two-mode squeezing transformation $\hat a_{\sigma,1}=\hat S^\dagger(r)\hat a_{\sigma,in} \hat S(r).$

As shown in Fig.~\ref{figure:QFI}, after passing through the sample each circular polarization mode of the TMSV undergoes a different gain. Using the input–output relation \eqref{ipop} together with Eq.~\eqref{a1tmsv}, the output operators are obtained as:
\begin{equation}
    \begin{aligned}
    \hat a_{R,2}=&\hat a_{R,in}\sqrt {G_0+G_\chi}\cosh{r}+\hat a_{L,in}^{\dagger}\sqrt {G_0+G_\chi}\sinh{r}\\
    &+\hat b^\dagger_{R}\sqrt {G_0+G_\chi-1}\\[2pt]
    \hat a_{L,2}=&\hat a_{L,in}\sqrt {G_0}\cosh{r}+\hat a_{R,in}^{\dagger}\sqrt {G_0}\sinh{r}\\
    &+\hat b^\dagger_{L}\sqrt {G_0-1},
\end{aligned}
\label{out}
\end{equation}

Since the displacement vector for the output state is zero for a TMSV probe undergoing gain, the QFI depends solely on the covariance matrix, which is found using Eq.~\eqref{out}:
\begin{equation}
    \begin{aligned}
    \sigma=\left(
    \begin{array}{cccc}
     \sigma _{11} & 0 & 0 & \sigma _{14} \\
     0 & \sigma _{22} & \sigma _{23} & 0 \\
     0 & \sigma _{32} & \sigma _{33} & 0 \\
     \sigma _{41} & 0 & 0 & \sigma _{44} \\
    \end{array}
    \right),
    \end{aligned}
    \label{covtmsv}
\end{equation}
where,  
\begin{align*}
    \sigma _{11}=&\sigma _{33}=( G_0+G_\chi) \cosh (2 r)+G_0+G_\chi-1\\
    \sigma _{22}=&\sigma _{44}=G_0 \cosh (2 r)+G_0-1\\ 
    \sigma _{41}=&\sigma _{32}=\sigma _{23}=\sigma _{14}=\sqrt{G_0(G_0+G_\chi)} \sinh (2 r),
\end{align*}
For the TMSV probe, the covariance matrix contains nonzero off-diagonal elements arising from the intermode correlations inherent to two-mode squeezing. These correlations become stronger with increasing squeezing parameter r, which can be achieved by using a stronger pump to generate the TMSV state.

Using the ordered basis $(G_0,G_\chi)$ the QFI matrix for the TMSV probe can be found using Eq.~\eqref{QFI} and Eq.~\eqref{covtmsv}. Refer to Appendix A for the full expression  of the QFI matrix. In a multi-parameter estimation the diagonal entries of the QFI matrix alone do not tell us how precisely the parameter can be estimated. The off-diagonal elements show that the estimation errors of the two parameters are correlated, so the uncertainty in one parameter influences the precision bound of the other. Therefore, one must consider the inverse of the QFI matrix, whose full expression is also given in the Appendix A. The diagonal elements of $\mathbf{F}_Q^{-1}$ determine the multiparameter quantum Cram\'er--Rao bounds and thus set the lower bounds on the variances of the parameter estimates achievable with the TMSV probe.

\begin{figure}
    \centering
    \begin{overpic}[width=1\linewidth]{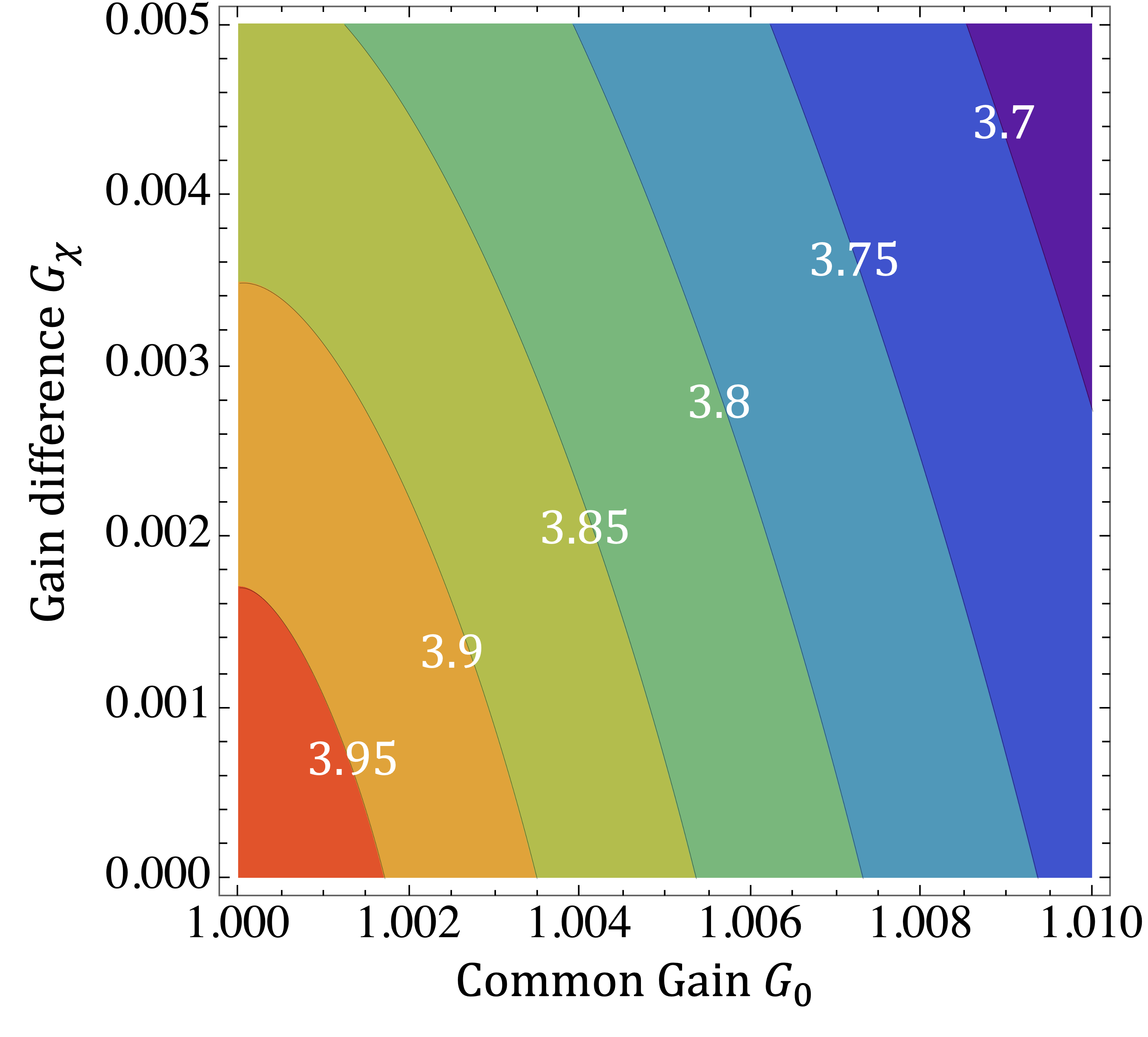}
    \end{overpic}
    \caption{
     Contour plot of Quantum Advantage (values shown in white)$=(\Delta G_\chi)^{\mathrm{Coh}}_{\mathrm{QCRB}}/(\Delta G_\chi)^{\mathrm{TMSV}}_{\mathrm{QCRB}}$  for mean photon number $\bar n=15$ in each circular polarization mode, as a function common gain $G_0$ and gain difference $G_\chi$. Values greater than unity indicate the enhancement in QCRB with the TMSV probe compared with a coherent probe.
    }
    \label{fig:QA}
\end{figure}

For a mean photon number of $\bar n=\sinh^2{r}$ in each mode of polarization, the QCRB for multiparameter estimation of the chiral gain difference $G_\chi$ is given by
\begin{equation}
    \begin{aligned}
    (\Delta G_\chi)^{\mathrm{TMSV}}_{\mathrm{QCRB}}& = \sqrt{\left(F_Q^{-1}\right)_{22}}\\
    \Rightarrow (\Delta G_\chi)^{\mathrm{TMSV}}_{\mathrm{QCRB}}&= \sqrt{\frac{G_\chi ^2+P}{\bar n+1}+\frac{\bar n R}{(\bar n+1) (P+1)}},
\end{aligned}
\label{TMSVqcrb}
\end{equation}

where,
\begin{align*}
    P&=2 G_0 (G_0+G_\chi +-1)-G_\chi\\
    R&=G_\chi^2 (G_0 - 1)(G_0+G_\chi - 1),
\end{align*}
A comparison of the QCRB for coherent state probe Eq.~\eqref{Cohqcrb} and TMSV probe Eq.~\eqref{TMSVqcrb} is used to determine the precision enhancement achievable using TMSV. This is done by defining the quantum advantage ($QA=(\Delta G_\chi)^{\mathrm{Coh}}_{\mathrm{QCRB}}/(\Delta G_\chi)^{\mathrm{TMSV}}_{\mathrm{QCRB}}$) as the ratio of QCRB for coherent state and TMSV. Fig.~\ref{fig:QA} shows a plot of QA as a function of $G_0$ and $G_\chi$. The mean photon number($\bar n$) in each mode of polarization the was kept the same ($\bar n=\alpha^2=\sinh^2{r}=15$) for both the coherent and the TMSV probes. The plot shows that the TMSV probe provides enhanced sensitivity over the coherent probe throughout the plotted region. And the quantum advantage becomes more prominent for smaller common gain($G_0$) and chiral gain difference($G_\chi$).  The plot can be understood from the process of Raman gain, which introduces additional uncorrelated noise to the field. Thus, in the low gain regime the TMSV probe retains stronger photon number correlation and enables more precise measurement over the coherent probe.

\begin{figure}
    \centering
    \begin{overpic}[width=1\linewidth]{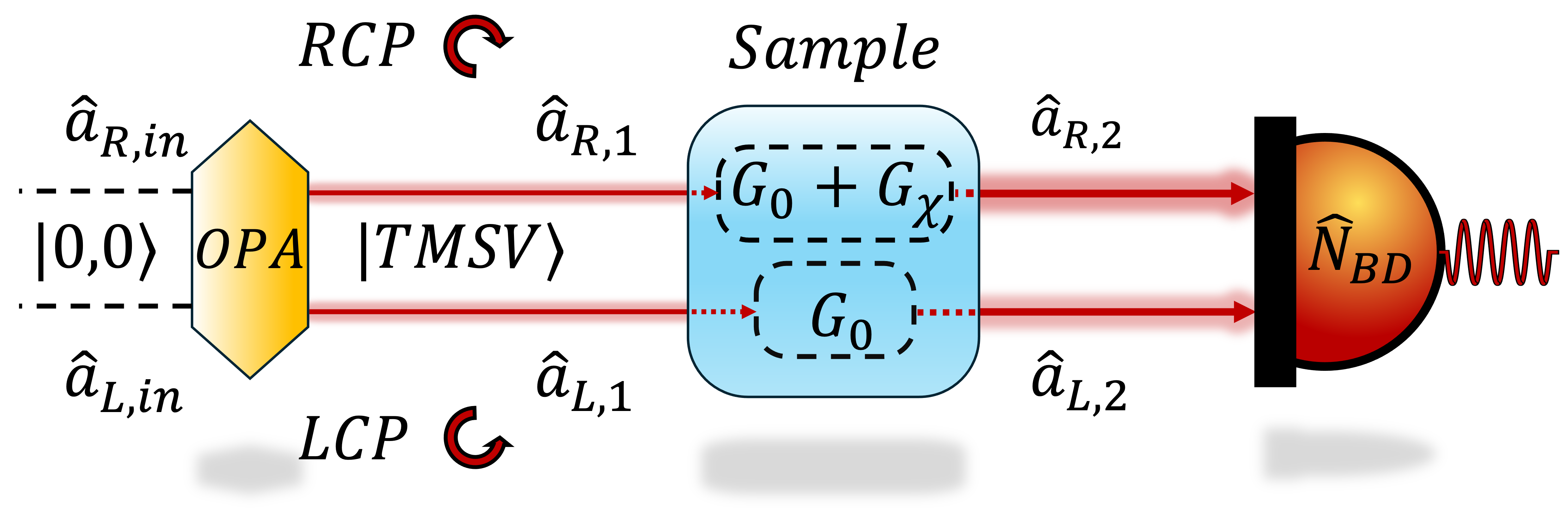}
    \end{overpic}
    \caption{Measuring the intensity difference between RCP and LCP modes of Stokes beam using balanced detection $\braket{\hat N_{\mathrm{BD}}}=\braket{\hat a_{R,2}^{\dagger}\hat a_{R,2}^{\vphantom{\dagger}}-\hat a_{L,2}^{\dagger}\hat a_{L,2}^{\vphantom{\dagger}}}$. Here the TMSV Stokes beam is produced using the OPA and the RCP and LCP modes undergoes gain $G_0+G_\chi$ and $G_0$ respectively, upon passing through the sample (The Pump used for the sample and OPA is not shown). The figure is same as Fig.~\ref{figure:QFI} with the addition of the detector.}
    \label{fig:BDmes}
\end{figure}

For many biosamples of interest $G_0-1$ is on the order of $10^{-4} \text{ to }10^{-7}$~\cite{steinle2016,floess2023,crisafi2017}. However, these gains can be enhanced by several order of magnitude using the resonant Raman process~\cite{oh26,ameer13}. In this process, the frequency of the pump field is chosen to be near the resonance of an electronic transition. This reduces the optical detuning and causes the effective Raman polarizability to increase, which amplifies the Stokes response. Thus, both the common gain $G_0$ and the chiral Raman gain $G_\chi$ becomes larger.
However, the gain difference between the RCP and LCP modes is much smaller than the common gain. And in the low gain difference limit, $|G_\chi|\ll( G_0-1)$ we get,
\begin{align}    (\Delta G_\chi)^{\mathrm{TMSV}}_{\mathrm{QCRB}}\approx\sqrt{\frac{2G_0(G_0-1)}{\bar n+1}}.
\label{TMSVQCRB}
\end{align}
Comparing this result with Eq.~\eqref{Cohqcrb}, we see that, to achieve the same sensitivity as a TMSV probe, a coherent probe requires $\gamma$ times more photons. Equivalently, $\bar n_{\mathrm{coh}}=\gamma \bar n_{\mathrm{sq}}$, where $\bar n_{\mathrm{coh}}$ and $\bar n_{\mathrm{sq}}$ are the average photon numbers required for the coherent and TMSV probes, respectively. For a common gain of $G_0-1=10^{-3}$ to achieve the same sensitivity as TMSV one would need a coherent probe that is around $10^3$ times more intense. Since many biological samples are damaged not only by high peak power but also by cumulative photon exposure, a TMSV probe reduces the total photon exposure and expands the applicability of ROA spectroscopy to a broader class of photosensitive biosamples.  

\section{The Optimal measurement scheme}

Having established the ultimate precision bound of $G_\chi$ measurement, we now identify a measurement scheme capable of approaching it. A joint estimation of $G_0$ and $G_\chi$ would require a joint photo detection of both polarizations modes. However, for the measurement of $G_\chi$ alone, it is sufficient to measure only the intensity difference between the two modes via balanced detection, since it can be deduced using the input and output photon number count using Eq.~\eqref{Nbd}. This scheme is shown in Fig.\ref{fig:BDmes}. Interestingly a balanced photodetection scheme achieves the same sensitivity as joint photodetection, through a single measurement, as shown in appendix B. Although the measurement of $G_\chi$ via balanced detection is not optimal in general, its error sensitivity approaches the QCRB in the limit of a small chiral gain difference, $|G_{\chi}|\ll G_{0}-1$, which is the regime relevant to stimulated ROA.

\begin{figure}
    \centering
    \begin{overpic}[width=1\linewidth]{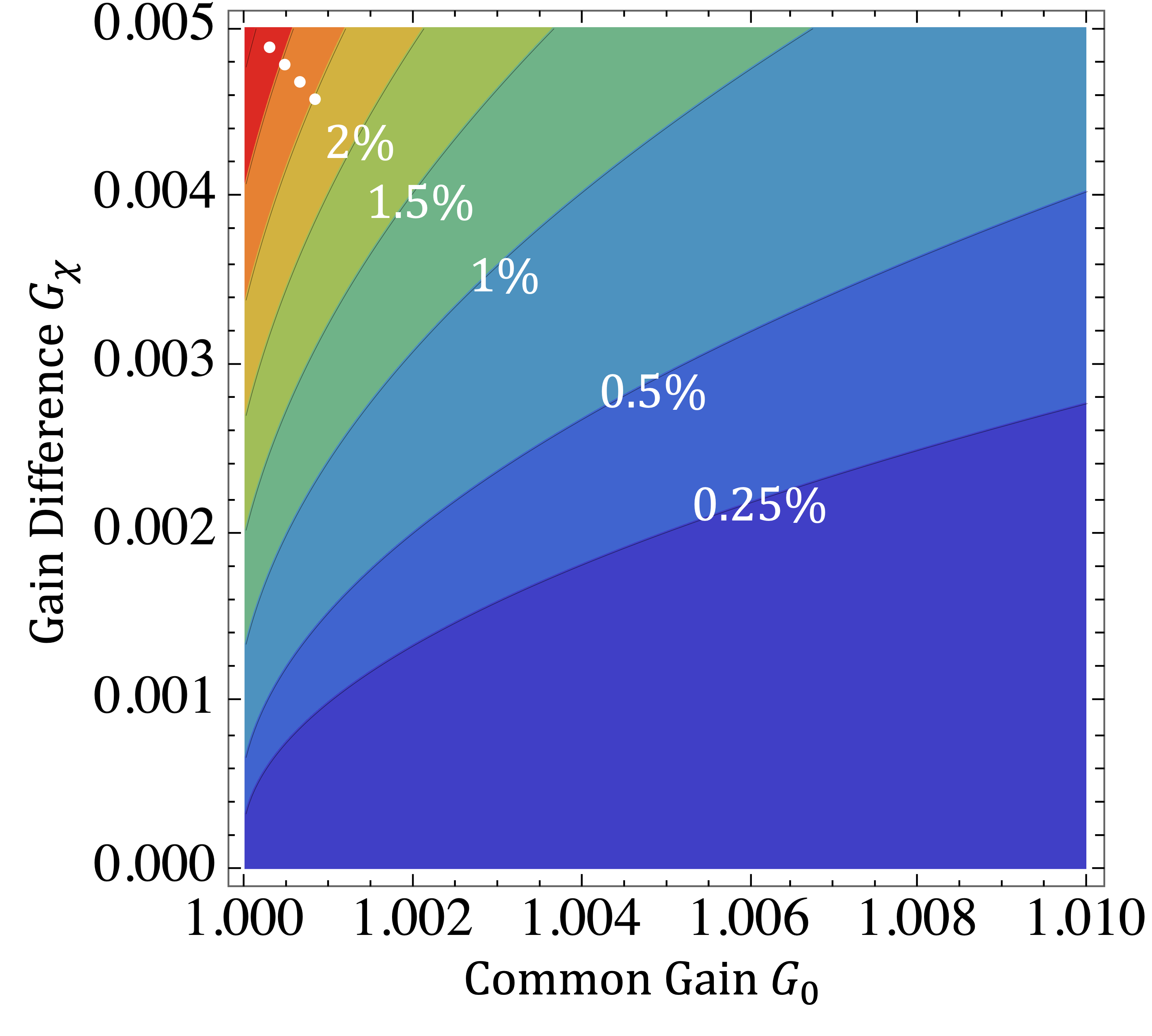}
    \end{overpic}
    \caption{Contour plot of normalized difference between the measured
    error sensitivity and QCRB $=((\Delta G_\chi)^{\mathrm{TMSV}}_{\mathrm{BD}}-(\Delta G_\chi)^{\mathrm{TMSV}}_{\mathrm{QCRB}})/(\Delta G_\chi)^{\mathrm{TMSV}}_{\mathrm{QCRB}}$ for TMSV probe. The white values show the percentage difference for a mean photon number of $\sinh^2{r}=\bar n=15$ in each circular polarization mode.}
    \label{fig:QFImes}
\end{figure}

For balanced detection, the signal is defined as the intensity difference of the two beams using the operator $\hat N_{\mathrm{BD}}=\hat a_{2,R}^{\dagger}\hat a_{2,R}^{\vphantom{\dagger}}-\hat a_{2,L}^{\dagger}\hat a_{2,L}^{\vphantom{\dagger}}$. And the input-output relations in Eq.~\eqref{out} can be used to find the expected photon number in the detector,
\begin{align}
\langle \hat N_{\mathrm{BD}}\rangle= (\bar n+1)G_\chi,
\label{Nbd}
\end{align}
and the variance of the signal,
\begin{align}
&\text{Var}(\hat N_{\mathrm{BD}}) =\braket {\hat N_{\mathrm{BD}}^2}-\braket {\hat N_{\mathrm{BD}}}^2\nonumber\\
&=(\bar n+1) \left(G_{\chi } \left((\bar n+1) G_{\chi }+2 G_0-1\right)+2 \left(G_0-1\right) G_0\right).
\end{align} 
Eq.\eqref{Nbd} can be used to find the sign and magnitude of the chiral gain difference $G_\chi$. The error in the estimation of $G_{\chi}$ is determined by the quantum fluctuation $\Delta N_{\mathrm{BD}}$ of the signal, arising from the intrinsic probabilistic nature of photodetection. We can find the error sensitivity of $G_{\chi}$ using,
\begin{equation}
\begin{aligned}
    \left(\Delta G_{\chi}\right)^{\mathrm{TMSV}}_{\mathrm{BD}}&=\frac{\sqrt{\text{Var}( \hat N_{\mathrm{BD}})}}{|d\langle \hat N_{\mathrm{BD}}\rangle/d G_\chi|}\\
    &=\sqrt{\left(\frac{\left(2 G_0-1\right) G_{\chi }+2 G_0\left(G_0-1\right) }{\bar n+1}+G_{\chi }^2\right)},
\end{aligned}
\label{errBD}
\end{equation}
The performance of balanced detection is shown in Fig.~\ref{fig:QFImes}, using the normalized difference (ND) between error sensitivity and the QCRB with TMSV,
\begin{align}
    \text{Normalized Difference}=\frac{(\Delta G_\chi)^{\mathrm{TMSV}}_{\mathrm{BD}}-(\Delta G_\chi)^{\mathrm{TMSV}}_{\mathrm{QCRB}}}{(\Delta G_\chi)^{\mathrm{TMSV}}_{\mathrm{QCRB}}},
\end{align}
 as a function of $G_0$ and $G_\chi$ for a mean photon number $\bar n=15$ in each circular polarization mode of light. The plot shows that balanced detection becomes more effective for smaller values of gain difference $G_\chi$. This measurement scheme is particularly effective for the high photon number correlated TMSV probe because taking the difference in intensity of the two polarization modes cancels the correlated noise and thus the signal noise at output is very small. For larger values of $G_\chi$ the imbalance in gain leaves larger residual noise in one arm, reducing the effectiveness of balanced detection. In the gain limit relevant to ROA ($|G_\chi|\ll G_0-1$) we have,
\begin{align} 
    \left(\Delta G_{\chi}\right)^{\mathrm{TMSV}}_{\mathrm{BD}}=\left(\Delta G_{\chi}\right)^{\mathrm{TMSV}}_{\mathrm{QCRB}}=\sqrt{\frac{2 G_0\left(G_0-1\right) }{\bar n+1}}.
    \label{BDsat}
\end{align}
This denotes the saturation of the QCRB for estimating $G_\chi$ only, and not for estimating the common gain $G_0$. 

\begin{figure}
    \centering
    \begin{overpic}[width=1\linewidth]{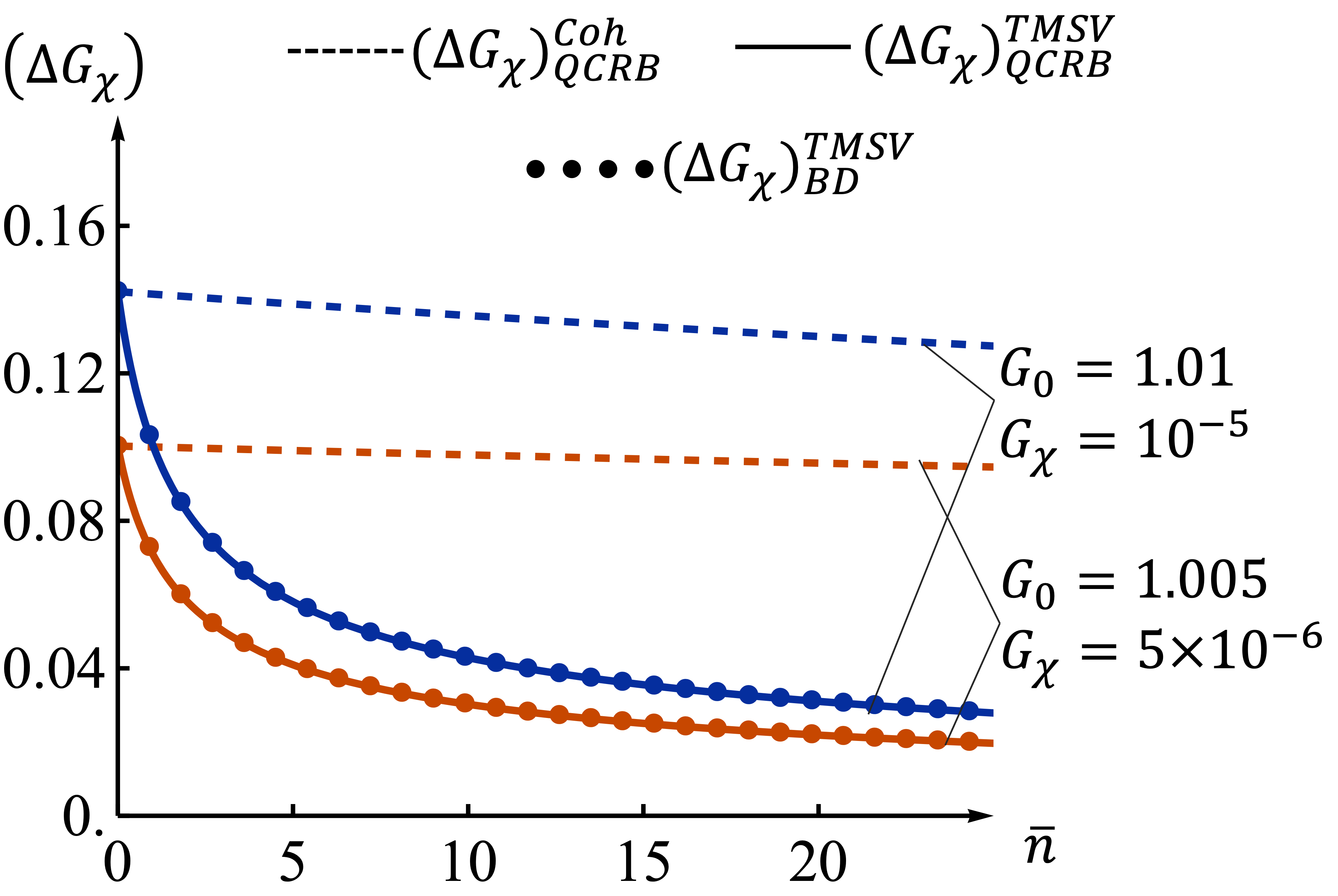}
    \end{overpic}
    \caption{The plot compares the $G_\chi$ estimation error from QCRB for a coherent probe (dashed curve), with the TMSV-probe QCRB (solid curves), and the measured error sensitivity for the TMSV probe(circular marker), for two sets of values of gains, $G_0=1.01,G_\chi=10^{-5}\text{ and }G_0=1.005,G_\chi=5\times10^{-6}$. The close overlap between the circular markers and the corresponding solid curves indicates that the measurement nearly saturates the QCRB.}
    \label{fig:Allplot}
\end{figure}

A comparison of the QCRB of $G_\chi$ estimation with coherent state, Eq.~\eqref{Cohqcrb}, with TMSV probe, Eq.~\eqref{TMSVqcrb} and error sensitivity for balanced detection with TMSV probe, Eq.\eqref{errBD}, is shown in Fig.~\ref{fig:Allplot}. It can be seen that the sensitivity achieved with TMSV probe is much higher than the coherent probe, and the difference increases steadily for greater mean photon number. This is expected since a TMSV probe with larger photon number allows for stronger correlation in photon number and it experiences a reduced intensity difference fluctuation caused by Raman gain process. Furthermore, the close overlap of circular markers with the corresponding solid lines shows that the balanced detection nearly saturates the QCRB. This is consistent with the finding in Eq.\eqref{BDsat}. The plot also shows that to achieve the sensitivity the TMSV probe much smaller photon number than the coherent probe.

\section{Robustness Against Detection Loss}

Photon loss is unavoidable in all optical measurements and can introduce additional vacuum noise to the measured field. It is particularly important for the TMSV probe because the observed precision enhancement depends on the photon number correlation, which can be degraded in the presence of photon loss. Thus, here we present our results on quantum advantage when detector losses are taken into account.

To model the loss, we considered a photodetector with finite quantum efficiency $\eta$. The detector efficiency for both modes are matched so that the intensity difference can be solely attributed to the chiral gain difference between the polarization modes. The field detected at the output can be described using the input-output relation 
\begin{equation}
\hat a_{\sigma,\mathrm{det}}
=
\sqrt{\eta}\,\hat a_{\sigma,2}
+
\sqrt{1-\eta}\,\hat \nu_{\sigma},
\qquad \sigma\in\{R,L\},
\label{ioloss}
\end{equation}
were $\hat a_{\sigma,\mathrm{det}}$ is the detected field operator for polarization $\sigma$, and $\hat a_{\sigma,2}$ is the corresponding output field after the sample. The parameter $\eta$ denotes the detector quantum efficiency, with $\eta=1$ for ideal detection. The operator $\hat \nu_{\sigma}$ is included to account for the noise introduced by the loss and it satisfies the relation $[\hat \nu_{\sigma},\hat \nu_{\sigma'}^\dagger]=\delta_{\sigma\sigma'}$. Using the Eq.\eqref{ioloss} one can find the final output operators and get the displacement and the covariance matrix. Following the same procedure as appendix A and B the QFI matrix and the error sensitivity in presence of the loss were obtained. 

\begin{figure}
    \centering
    \begin{overpic}[width=1\linewidth]{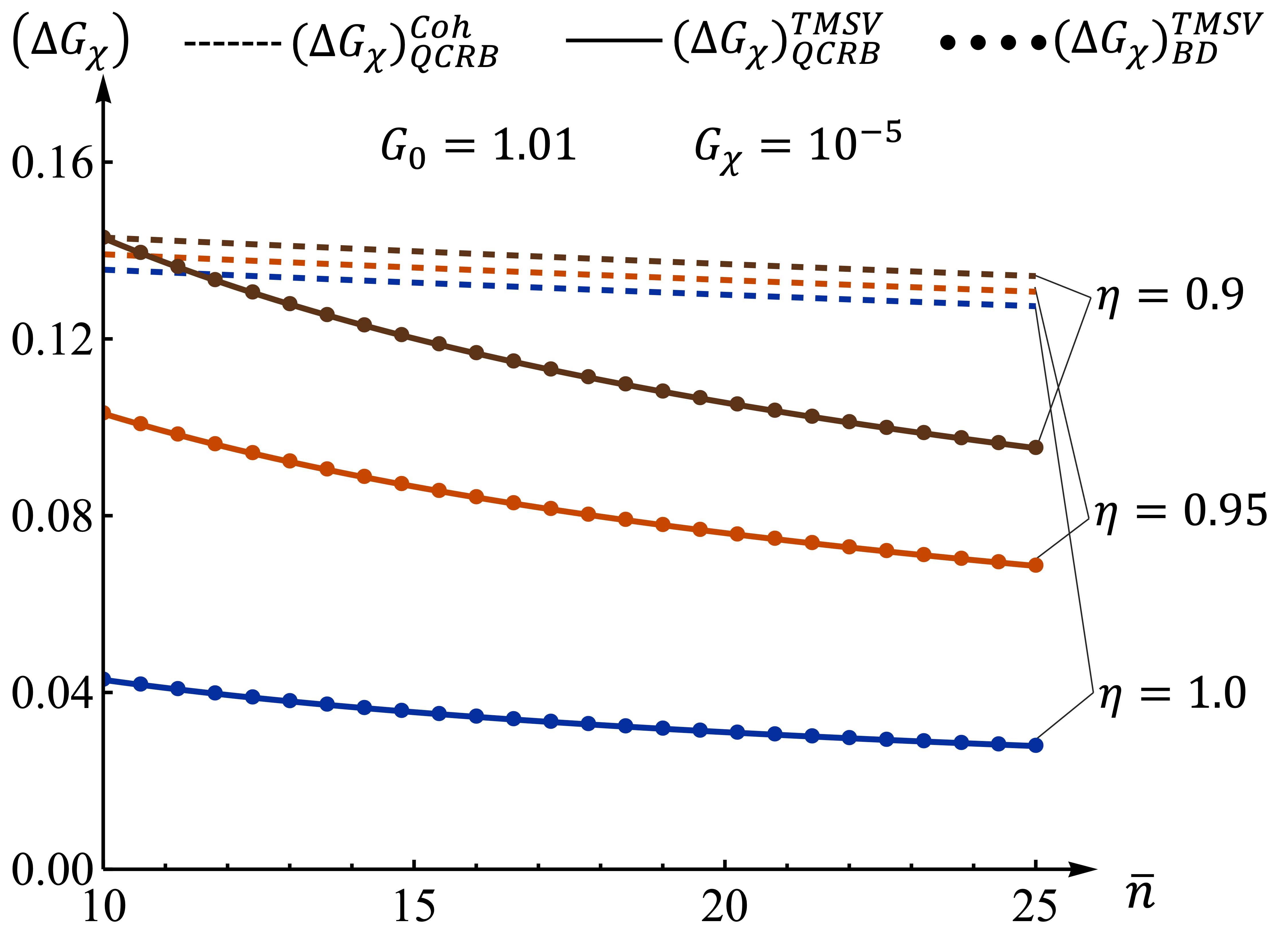}
    \end{overpic}
    \caption{Uncertainty bound for $G_\chi$ measurement against mean photon number($\bar n$) has been plotted for several values of detector efficiency($\eta=0.9,0.95 \text{ and } 1$). The QCRB of $G_\chi$ estimation with coherent probe and TMSV probe has been represented with dashed and solid curves respectively. The circular markers represent the error sensitivity obtained with balanced detection, and their close overlap with the corresponding solid curves shows its near optimal performance. The chosen photon number range is when quantum advantage occurs.}
    \label{fig:errloss}
\end{figure}

The quantum efficiency of many modern detectors could be in the range of $90\%$ or above~\cite{mehmet11,muramatsu97}, thus we have evaluated the effect of finite detection efficiency($\eta$) in this range. Fig.\ref{fig:errloss} shows how the $G_\chi$ measurement sensitivity compares with the coherent probe for non-ideal detectors. For all values of $\eta$ the precision enhancement with TMSV is found to be more prominent at larger mean photon number. This can be attributed to the stronger photon number correlation for the TMSV probe. However, a finite detection efficiency introduced additional uncorrelated vacuum noise to the probe state causing the the uncertainty of the TMSV probe to increase. Despite this the TMSV can be seen to offer significantly improved sensitivity over the coherent probe. Furthermore the circular markers lying on the solid lines shows near optimality of the balanced detection even in the presence of the photon loss.

\section{Discussion}
We calculated the Quantum Fisher information matrix to find the QCRB for gain difference estimation in stimulated Raman optical activity using a two-mode squeezed vacuum(TMSV) probe and compared it with that of a classical probe. For a fixed photon number in each probe, a better precision bound was shown to be achievable with TMSV in the low gain regime, which is relevant to stimulated ROA measurement. This enhancement is due to photon number correlation, and in the weak gain regime the correlation is less degraded, thus the noise in intensity difference remains low. Thus, the quantum advantage does not only come from increasing the signal itself but also by reducing the intrinsic quantum noise.  

Although multiparameter estimation often suggests the need for joint measurements on multiple output modes, here the chiral gain difference can be inferred from a single balanced detection measurement of the output intensity difference. This experimental simplicity is important in practice, since balanced detection is significantly easier to implement than joint photodetection. Our comparison with the quantum Cram\'{e}r--Rao bound shows that balanced detection attains the same sensitivity as joint photodetection and becomes asymptotically optimal in the limit $|G_\chi|\ll G_0-1$, where it saturates the QCRB. This bound represents an ideal theoretical limit. In practice, finite detector efficiency degrades the achievable sensitivity. However, the precision enhancement was shown to remain significant even in the presence of loss.

These results are promising for the use of ROA measurements on aqueous biosamples and drug molecules where increasing the probe power is not an option due to effects such as photodamage and photobleaching ~\cite{li2021srs,jones2019,zhang2022phototoxic,blanch2001}. Thus, TMSV provides a route to achieve enhanced sensitivity by lowering the error sensitivity at a fixed optical intensity. The combination of quantum-enhanced precision and experimentally accessible balanced detection therefore makes this scheme attractive for ROA spectroscopy in situations where the signal is weak and the allowable optical power is limited.

\section*{Acknowledgement}

The author is grateful for support through NSF Award No.
2426699 and the Robert A. Welch Foundation Grant
No. A-1943-20240404

\section*{Data Availability}
No data were created or analyzed in this study.

\onecolumngrid
\appendix
\setcounter{equation}{0}
\renewcommand{\theequation}{A\arabic{equation}}
\section*{Appendix A: The Quantum Cram\'{e}r--Rao bound for TMSV}

In this appendix we outline the derivation of the quantum Cram\'{e}r--Rao bound (QCRB) for a two-mode squeezed vacuum (TMSV) probe. Because the TMSV state is a zero-mean Gaussian state, all information about the estimation problem is contained in its covariance matrix. We consider the simultaneous estimation of the common Raman gain $G_0$ and the chiral gain difference $G_\chi$, which are the two parameters relevant to stimulated Raman optical activity in our model.

Starting from the output covariance matrix in Eq.~(18), we evaluate the quantum Fisher information matrix (QFIM) in the parameter basis $(G_0,G_\chi)$ using the Gaussian-state formula given in Eq.~(5). The resulting QFIM quantifies the ultimate sensitivity allowed by quantum mechanics for unbiased estimation of these two parameters. The QFI is given by,

\begin{align}
\mathbf{F}_{Q}
=
\dfrac{\bar n+1}{P(\bar n+1)+1}\begin{pmatrix}
\dfrac{\bar n G_\chi ^2}{G_0 (G_0+G_\chi)}+\dfrac{(P+1) \left(P+G_\chi ^2\right)}{UV} & \dfrac{P+1}{U}-\dfrac{G_\chi  \bar n}{G_0+G_\chi }\\
\dfrac{P+1}{U}-\dfrac{G_\chi  \bar n}{G_0+G_\chi} & \dfrac{P+1}{U}+\dfrac{G_0 \bar n}{G_0+G_\chi}
\end{pmatrix}.
\end{align}
Its matrix inverse then gives the corresponding multiparameter QCRB, from which the minimum variance bound for estimating $G_\chi$ is obtained from the $(2,2)$ element,
\begin{align}
\mathbf{F}_Q^{-1}
=
\frac{(G_0-1) }{\bar n+1}\left(
\begin{array}{cc}
 G_0 \left(\dfrac{\bar n (P+G_\chi )}{2 (P+1)}+1\right) & \dfrac{G_\chi  \bar n (P+G_\chi)}{2 (P+1)}-G_0 \\
 \dfrac{G_\chi  \bar n (P+G_\chi )}{2 (P+1)}-G_0 & \dfrac{\bar n R+(P+1) \left(P+G_\chi ^2\right)}{(G_0-1) (P+1)} \\
\end{array}
\right),
\end{align}
where,
\begin{align}
&P=2 G_0 (G_0+G_\chi-1)-G_\chi\nonumber,\\
&R=G_\chi^2 (G_0 - 1)(G_0+G_\chi - 1)\nonumber,\\
&U=(G_0+G_\chi-1) (G_0+G_\chi)\nonumber,\\
&V=G_0 (G_0 - 1).\nonumber
\end{align}
Therefore, the quantum Cram\'{e}r--Rao bound associated with estimating the chiral gain difference $G_\chi$ is obtained from the $(2,2)$ element of $\mathbf{F}_Q^{-1}$. This gives the fundamental precision bound for the TMSV probe in the present two-parameter estimation problem Eq.~\eqref{TMSVqcrb},
\begin{align}
    (\Delta G_\chi)^{TMSV}_{QCRB}=\sqrt{\frac{P+G_\chi ^2}{\bar n+1}+\frac{\bar n R}{(\bar n+1) (P+1)}}.
\end{align}

\setcounter{equation}{0}
\renewcommand{\theequation}{B\arabic{equation}}
\section*{Appendix B: Error sensitivity with Joint Photodetection }

In this appendix we derive the error sensitivity for estimating the chiral gain difference $G_\chi$ using joint photodetection. Since both the common gain $G_0$ and the chiral gain difference $G_\chi$ are treated as unknown parameters, the appropriate error-propagation formula is the covariance-matrix form of multiparameter estimation. Joint photodetection retains the full two-output covariance information and therefore provides a natural benchmark measurement for the two-parameter estimation problem.
We consider the photon-number observables $
\hat N_R=\hat a_{R,2}^\dagger \hat a^{\vphantom{\dagger}}_{R,2}\text{ and }
\hat N_L=\hat a_{L,2}^\dagger \hat a^{\vphantom{\dagger}}_{L,2}$ for the right- and left-circularly polarized output modes, respectively. Their mean values are
\begin{align}
\langle \hat N_L\rangle &= (\bar n+1)G_0-1,\\
\langle \hat N_R\rangle &= (\bar n+1)(G_0+G_\chi)-1.
\end{align}
For a single parameter $\theta$ estimated from an observable $\hat N$, the usual error sensitivity
$\Delta \theta$ is related to the estimator variance by $(\Delta \theta)^2 = \mathrm{Var}(\hat\theta)$ and linear error propagation gives
\begin{align}
\mathrm{Var}(\hat N)
=
(\Delta\theta)^2
\left(
\frac{\partial \langle \hat N\rangle}{\partial \theta}
\right)^2.
\end{align}

Error propagation for multiple parameters from a correlated observable is discussed in chapter 2.6 of Ref.~\cite{clifford1973}. For the joint estimation of $G_0$ and $G_\chi$, let $\hat{G}_0$ and $\hat{G}_\chi$ denote unbiased estimators of the parameters $G_0$ and $G_\chi$ respectively. The corresponding error-propagation formula then generalizes to covariance-matrix form. To express the multiparameter error-propagation formula in a compact form, we introduce vector notation for the observables and the estimators, along with their corresponding covariance matrices.
\begin{align}
\hat{\mathbf N}
&\equiv
\begin{pmatrix}
\hat N_L\\
\hat N_R
\end{pmatrix},
\qquad
\hat{\mathbf G}
\equiv
\begin{pmatrix}
\hat G_0\\
\hat G_\chi
\end{pmatrix},
\\[4pt]
\boldsymbol{\sigma}_N
&\equiv
\mathrm{Cov}(\hat{\mathbf N},\hat{\mathbf N})
=
\begin{pmatrix}
\mathrm{Var}(\hat N_L) & \mathrm{Cov}(\hat N_L,\hat N_R)\\
\mathrm{Cov}(\hat N_R,\hat N_L) & \mathrm{Var}(\hat N_R)
\end{pmatrix},
\\[4pt]
\boldsymbol{\sigma}_G
&\equiv
\mathrm{Cov}(\hat{\mathbf G},\hat{\mathbf G})
=
\begin{pmatrix}
\mathrm{Var}(\hat G_0) & \mathrm{Cov}(\hat G_0,\hat G_\chi)\\
\mathrm{Cov}(\hat G_\chi,\hat G_0) & \mathrm{Var}(\hat G_\chi)
\end{pmatrix}.
\end{align}
Since the mean photon counts depend linearly on $G_0$ and $G_\chi$, relation between the covariance of measurement and the estimators is determined by the constant Jacobian,

\begin{align}
\boldsymbol{\sigma}_N = J \boldsymbol{\sigma}_G J^T,
\qquad
J =
\begin{pmatrix}
\dfrac{\partial \langle \hat N_L\rangle}{\partial G_0} &
\dfrac{\partial \langle \hat N_L\rangle}{\partial G_\chi} \\[6pt]
\dfrac{\partial \langle \hat N_R\rangle}{\partial G_0} &
\dfrac{\partial \langle \hat N_R\rangle}{\partial G_\chi}
\end{pmatrix}
=
(\bar n+1)
\begin{pmatrix}
1 & 0\\
1 & 1
\end{pmatrix}.
\end{align}
Using the input-output relations in Eq.~{16} and Eq.~{1}, the covariance matrix is for $\boldsymbol{\hat N}$ is found to be
\begin{align}
\boldsymbol{\sigma}_N
=
(1 + \bar n) \begin{pmatrix}
G_0 \left(G_0 (\bar n+1)-1\right) & G_0 \bar n \left(G_0+G_{\chi }\right)\\
G_0 \bar n \left(G_0+G_{\chi }\right) &
\left(G_0+G_{\chi }\right) \left((\bar n+1) (G_0+G_{\chi })-1\right)
\end{pmatrix}.
\end{align}
The off-diagonal elements of $\boldsymbol{\sigma}_N$ represent the correlations between the detected photon numbers in the two output modes, which arise from the photon-number correlations of the TMSV probe after amplification. We also find inverse of the Jacobian,
\begin{align}
J^{-1}
=
\frac{1}{\bar n+1}
\begin{pmatrix}
1 & 0\\
-1 & 1
\end{pmatrix}.
\end{align}
Therefore, the covariance matrix of the estimators is obtained as
\begin{align}
\boldsymbol{\sigma}_G
=
J^{-1}\boldsymbol{\sigma}_N (J^{-1})^T
=
\frac{1}{\bar n+1}\left(
\begin{array}{cc}
 G_0 \left(G_0 (\bar n+1)-1\right) & G_0 \left(\bar n G_{\chi }-G_0+1\right) \\
 G_0 \left(\bar n G_{\chi }-G_0+1\right) & \left(2 G_0-1\right) G_{\chi }+2 G_0\left(G_0-1\right)+(\bar n+1)G_{\chi }^2 \\
\end{array}
\right).
\end{align}
Since $G_\chi$ is the second parameter in the estimator vector $\hat{\mathbf G}$, its estimation variance is given by the $(2,2)$ element of this matrix, $\Delta G_\chi=\sqrt{(\boldsymbol{\sigma}_G)_{22}}=\sqrt{\left(\frac{\left(2 G_0-1\right) G_{\chi }+2 G_0\left(G_0-1\right) }{\bar n+1}+G_{\chi }^2\right)}$, which agrees with the bound Eq.~\eqref{errBD} obtained via the balanced detection.

\setcounter{equation}{0}
\renewcommand{\theequation}{C\arabic{equation}}
\section*{Appendix C: Deriving input-output relation Eq.(1)}
We present a simple argument to bring out the physical meaning of noise terms in the basic equations Eq.\eqref{ipop}. We keep the analysis simple by ignoring any chiral effects. Let $a \text{ and } v$ be the annihilation opeartors associated with the stimulating Raman field and the vibrational mode. The effective Hamiltonian which describes the Raman process can be written in the form
\begin{equation}
    \mathcal{H} = g\epsilon_la^\dagger v^\dagger \text{exp}\{-i\delta t\}+H.C,
\end{equation}
where g is the Raman coupling, $\epsilon_l$ is the amplitude of the pumping field and $\delta=(\omega_l-\omega_s-\omega_R)$ with $\omega_l$ as pump frquency, $\omega_s$ Stokes frequency and $\omega_R$ frequency of Raman transition. For Resonant case $\delta=0$ and for simplicity we use it. The equations of motin for the annihilation operators $ a \text{ and } v$ are given by,
\begin{equation}
    \begin{aligned}
        \dot a &=-i g\epsilon_lv^\dagger, \\
        \dot v &=-ig\epsilon_la^\dagger-\frac{v}{T_2}+f_v(t).
    \end{aligned}
\end{equation}
Here $T_2$ is the lifetime of Raman mode and $f_v$ is the vibrational noise term. Note that 
\begin{equation}
    \begin{aligned}
        \braket{f_v(t)f_v^\dagger(t'}&=\frac{2}{T_2} (n_0+1)\delta(t-t'),\\
        \braket{f_v^\dagger(t)f_v(t'))}&= \frac{2}{T_2} (n_0)\delta(t-t').
    \end{aligned}
\end{equation}
Here $n_0$ is the number of thermal phonons which are negligible for room temperature and optical phonons. These relations are well known for any damped system [Eq(9.86),ref.~\cite{Agarwal_2012}] .
For most Raman trnsition $T_2$ is very short and hence we can drop $\dot v$ and solve for $v$
\begin{equation}
    v=-ig\epsilon_la^\dagger T_2+T_2f_v(t).
    \label{nuout}
\end{equation}
We use result Eq.\eqref{nuout} in the equation for the annihilation operator a to obtain,
\begin{equation}
    \dot a =+|g|^2|\epsilon_l|T_2a-ig\epsilon_lT_2f_v^\dagger(t).
    \label{aout}
\end{equation}
Solution of Eq.\eqref{aout} is
\begin{align*}
    a(t)=\text{exp}\{|g|^2T_2|\epsilon_l|^2t\}a(0)-ig\epsilon_lT_2\int_0^te^{|g|^2T_2|\epsilon_l|^2\tau}f^\dagger_v(t-\tau)d\tau,
\end{align*}
which can be written in the form
\begin{align}
    a(t)=\sqrt{G}a(0)+\sqrt {(G-1)}b^\dagger,\ \ \ [b,b^\dagger]=1,
\end{align}
and b is the annihilation operator associated with the vibrational noise, which is in vacuum state. Here we have defined $G=exp\{|g|^2T_2|\epsilon_l|^2\tau\}$ as the Raman gain. The equation for the Stokes field has the same form as Eq.\eqref{ipop}.

\bibliographystyle{ieeetr}
\bibliography{references.bib}

\end{document}